\documentclass[12pt]{article}
\usepackage{amssymb}
\usepackage{graphicx}
\usepackage{indentfirst}
\usepackage{cite}

\begin{document}

\title{Relation between quark-antiquark potential \\
and quark-antiquark free energy in hadronic matter}
\author{Zhen-Yu Shen and Xiao-Ming Xu}
\date{}
\maketitle \vspace{-1cm}
\centerline{Department of Physics, Shanghai University,
Baoshan, Shanghai 200444, China}

\begin{abstract}
In the high-temperature quark-gluon plasma and its subsequent
hadronic matter created in a high-energy nucleus-nucleus collision, the
quark-antiquark potential depends on the temperature. The temperature-dependent
potential is expected to be derived from the free energy obtained in lattice
gauge theory calculations. This requires one to study the relation between the 
quark-antiquark potential and the quark-antiquark free energy.
When the system's temperature is above the critical temperature, the potential
of a heavy quark and a heavy antiquark almost equals the free energy, but the
potential of a light quark and a light antiquark, of a heavy quark and a light
antiquark and of a light quark and a heavy antiquark
is substantially larger than the free energy. When the system's temperature
is below the critical temperature, the quark-antiquark free energy can be taken
as the quark-antiquark potential. This allows one to apply the quark-antiquark
free energy to study hadron properties and hadron-hadron reactions in hadronic
matter.
\end{abstract}

\noindent
PACS: 25.75.-q; 25.75.Nq; 12.38.Mh

\noindent
Keywords: Potential; Free energy; Hadronic matter.

\section{Introduction}
The quark-antiquark free energy $F$ is defined as the quark-antiquark internal
energy $U$ minus the product of the temperature $T$ and the quark-antiquark
entropy $S$. The
internal energy of a quark and an antiquark at rest  is the quark-antiquark
potential. When the temperature is above the critical temperature $T_{\rm c}$,
the quark-antiquark free energy can not be identified as the quark-antiquark
potential \cite{wong2}. 
From a correlator of a very heavy quark-antiquark pair a time-dependent
potential was obtained in quenched lattice QCD \cite{RHS,BR}. Only at very 
large times and at $T<T_{\rm c}$ the potential agrees with the free energy in 
the Coulomb gauge. At $T>T_{\rm c}$ the potential at any time
deviates from the free energy. In determining the potential 
there is uncertainty from the form of the correlator. The potential at very
large times is thought to be the quark-antiquark potential.
Hence, the relation between the quark-antiquark potential and the
quark-antiquark free energy is not well understood. The
medium created in high-energy nucleus-nucleus collisions is hadronic matter and
the quark-gluon plasma. On the one hand hadron properties and hadron-hadron
reactions change considerably from vacuum to hadronic matter with the
change of the quark potential \cite{ZXG,ZX}, on the other
hand the most striking medium effect is the QCD phase transition
in which quark behavior and quark confinement change \cite{satz}. 
Hence, we need the change of the quark
potential with respect to vacuum, hadronic matter, and the quark-gluon plasma,
i.e. the temperature dependence of the quark potential. At present, the 
temperature-dependent potential is expected to be derived from the free energy
obtained in lattice gauge theory calculations. Therefore, we must study the
relation between the quark-antiquark potential and the quark-antiquark 
free energy.

In thermodynamics the entropy indicates the disorder of random motion
of particles in a system that consists of a large number of particles.
It equals the negative of the derivative of the 
system free energy with respect to the temperature
while the system volume $V$ is fixed. While $V$ is fixed, the
distance $r$ between any quark and any antiquark always changes.
Then, the quark-antiquark entropy cannot be taken as the negative of the
derivative of the quark-antiquark free energy $F(T,r)$ with respect to the
temperature while $r$ is fixed. How to get the quark-antiquark entropy from
the quark-antiquark free energy is a problem in lattice QCD. A
formula for calculating the entropy from the QCD partition function
with additional heavy quarks is given in Ref. \cite{KZ}. The partition function
contains the QCD action and an operator that is proportional to
the product of two thermal Wilson lines. One needs to calculate the expectation
values of the operator, of the derivative of the action with respect to the
temperature and of the product of the operator and the derivative of the 
action with respect to the temperature. In addition, a constant needs to be 
fixed through renormalization.
The formula is thus complicated and, in particular, not suitable for
lattice simulations.

We must find another way to calculate the quark-antiquark entropy. The entropy
is the thermodynamic quantity that characterizes the disorder of thermal
motion of particles in a system. It is obtained from observables
in thermodynamics. In the present work we evaluate the
quark-antiquark entropy from energy densities and pressure of particle systems.
A quark-gluon plasma created in a high-energy nucleus-nucleus collision expands
while its temperature decreases, and the plasma 
hadronizes at the critical temperature to produce hadronic matter. 
Hadronic matter expands while its temperature decreases
toward the freeze-out temperature. Quarks and antiquarks in the quark-gluon
plasma and pions that dominate hadronic matter are the particle systems 
what we concern.

\section{Free energy and potential}
The quark-antiquark internal energy is related to the quark-antiquark free
 energy by
\begin{equation}
U (T, r) =F (T, r)+TS,
\end{equation}
where $r$ is the distance between the quark and the antiquark.
The free energy is given by the Polyakov loop correlation function in lattice
QCD. If $TS$ is small, the quark-antiquark free energy can be considered as the
quark-antiquark potential. If the free energy is of a heavy quark and a heavy
antiquark, we calculate the entropy of this heavy quark-antiquark pair. If the
free energy is of a light quark and a light antiquark, we calculate the entropy
of this light quark-antiquark pair.
In the following two subsections we show that $TS$ is generally
comparable to the free energy in a 
quark-gluon plasma, but is small in hadronic matter.

The quark-gluon plasma and hadronic matter contain a multitude of 
quark-antiquark pairs. Different quark-antiquark pairs may have different free
energies, and the free energies depend on gluon states that propagate between
the quark and the antiquark \cite{MS,rothe}. The free energy is an extensive
variable, and the quark-antiquark free energy is the system's free energy
divided by the number of quark-antiquark pairs.

\subsection{In the case of massless particles}
In a quark-gluon plasma with three massless flavors, the quark energy density
is \cite{wong3}
\begin{equation}
\epsilon_q=g_Q\frac {7\pi^2}{240}T^4,
\end{equation}
with the color-spin-flavor degeneracy factor $g_Q=18$, and the antiquark
energy density is
\begin{equation}
\epsilon_{\bar q}=g_{\bar Q}\frac {7\pi^2}{240}T^4,
\end{equation}
with $g_{\bar Q}=18$. The pressure due to quarks is
$P_q=\frac {1}{3} \epsilon_q$,
and the pressure due to antiquarks is
$P_{\bar q}=\frac {1}{3} \epsilon_{\bar q}$.
The quark entropy density is
\begin{equation}
s_q=\frac {\epsilon_q + P_q}{T}=\frac {4}{3} \frac {\epsilon_q}{T},
\end{equation}
and the antiquark entropy density is
\begin{equation}
s_{\bar q}=\frac {\epsilon_{\bar q} + P_{\bar q}}{T}
=\frac {4}{3} \frac {\epsilon_{\bar q}}{T}.
\end{equation}
The quark number density is \cite{wong3}
\begin{equation}
n_q=g_Q \frac{3\zeta(3)}{4\pi^2 } T^3,
\end{equation}
with $\zeta(3)=1.20205$, and the antiquark number density is 
\begin{equation}
n_{\bar q}=g_{\bar Q}\frac{3\zeta(3)}{4\pi^2 } T^3.
\end{equation}

\indent There are one quark and one antiquark in the volume $V_{\rm c}=1/n_q
=1/n_{\bar q}$. The dimensionless entropy of the pair of quark and antiquark is
proportional to
\begin{equation}
S_{q\bar q}=s_qV_{\rm c}+s_{\bar q}V_{\rm c}
=\frac{14\pi^4}{135\zeta(3)}\simeq 8.408,
\end{equation}
which is independent of temperature. On the average the spatial separation of
the quark and the antiquark is
\begin{equation}
\sqrt[3]{V_{\rm c}}=\frac{1}{T}\sqrt[3]{\frac{4\pi^2}{3g_Q\zeta(3)}}.
\end{equation}

At $T=1.01T_{\rm c}$ with the critical
temperature $T_{\rm c}=0.175$ GeV \cite{KLP,GLMRX}, the spatial separation
is 0.946 fm where the free energy given in the lattice calculations is about
0.123 GeV \cite{KLP}. The free energy is only for a pair of
quark and antiquark in a flavor and in the color singlet. 
However, the quark (antiquark) entropy density $s_q$ ($s_{\bar q}$) used in 
Eq. (8) corresponds to the three flavors and the three colors. 
$\frac {1}{3} s_q$ 
($\frac {1}{3} s_{\bar q}$) is the quark (antiquark) entropy density for a
flavor or for a color. Hence, $\frac {1}{3}$ is multiplied to
$S_{q\bar q}$ to get the entropy of the pair of quark
and antiquark in a flavor, and another $\frac {1}{3}$ is multiplied to
$S_{q\bar q}$ to get the entropy of the pair of quark
and antiquark in the color singlet. $\frac {1}{3}\frac {1}{3} S_{q\bar q}$ is
the entropy of the pair of quark and antiquark in a flavor and 
in the color singlet. $\frac {1}{3}\frac {1}{3} TS_{q\bar q}$ is
then compared to the free energy given in the lattice calculations.
$\frac {1}{3}\frac {1}{3} TS_{q\bar q}$
takes the value 0.165 GeV and is larger than the free energy 0.123 GeV. 
The quark-antiquark internal energy is by 0.165 GeV larger than the free 
energy. By subtracting the total kinetic energy of the quark and the antiquark,
which is 0.062 GeV in the volume $V_{\rm c}$, we get the quark-antiquark
potential which is by 0.103 GeV larger than the free energy.
In the evolution of the quark-gluon
plasma entropy is conserved. When temperature increases from 
$T_{\rm c}$, the free energy decreases \cite{KLP}, the ratio of
$\frac {1}{3}\frac {1}{3}TS_{q\bar q}$ to the free energy increases,
and the difference between the quark-antiquark potential and the free energy
becomes larger and larger.
Therefore, the quark-antiquark free energy can not be taken as the
quark-antiquark potential in the quark-gluon plasma in the case of massless
quarks and antiquarks.

The energy density of a system of massless pions is \cite{wong3}
\begin{equation}
\epsilon_{\pi}=g_{\pi}\frac {\pi^2}{30}T^4,
\end{equation}
with $g_{\pi}=3$ for $\pi^+$, $\pi^{\rm 0}$, and $\pi^-$.
The pressure due to massless pions is $P_{\pi}=\frac {1}{3} \epsilon_{\pi}$.
The pion entropy density is
\begin{equation}
s_{\pi}=\frac {\epsilon_{\pi} + P_{\pi}}{T}=\frac {4}{3}
\frac {\epsilon_{\pi}}{T}.
\end{equation}

At $T_{\rm c}$ the quark and the antiquark in the volume $V_{\rm c}$ transit
into a pion. The dimensionless pion entropy in this volume is proportional to
\begin{equation}
S_{\pi}=s_{\pi}V_{\rm c}=\frac{8\pi^4 g_{\pi}}{135g_{Q}\zeta(3)} \approx 0.8,
\end{equation}
which is independent of temperature.
The entropy that corresponds to the quark-antiquark free
energy is $\frac {1}{3}S_{\pi}$ for one isospin component. 
The free energy slightly below $T_{\rm c}$ is about 0.29 GeV \cite{KLP}.
The ratio of $\frac {1}{3}TS_{\pi}= 0.046$ GeV 
with $T=0.99T_{\rm c}$ to the free energy is about 0.16. 
The quark-antiquark internal energy is by 0.046 GeV larger than the free 
energy. By subtracting the total kinetic energy of the quark and the antiquark,
which is 0.0128 GeV in the center-of-mass frame of the pion, we get the 
quark-antiquark potential which is by 0.0332 GeV larger than the free energy.
While hadronic matter expands, entropy is conserved,
temperature decreases, the free energy increases \cite{KLP}, the ratio of
$\frac{1}{3}TS_{\pi}$ to the free energy decreases, and the difference between
the quark-antiquark potential and the free energy becomes smaller and smaller.
Hence, we conclude that $U(T,r) \approx F(T,r)$ in hadronic matter and the free
energy can be taken as the quark-antiquark potential.

\subsection{In the case of massive particles}
The grand partition function for a quark system is
\begin{equation}
\Xi = \prod\limits_l (1+e^{\beta (\mu - \varepsilon_l)})^{\omega_l}, 
\end{equation}
where $\beta=1/T$, $\mu$ is the quark chemical potential,
and $\omega_l$ is the number of states corresponding to the 
quark energy $\varepsilon_l$. In the relativistic case a quark has the
energy $\varepsilon_l=\sqrt {\vec {p}^2 +m^2}$. Quark, antiquark and gluon
distributions in the quark-gluon plasma are described by the J\"uttner
distribution with the fugacity $\lambda=\exp (\mu/T)$. The evolution of the
quark-gluon plasma
can be determined by a set of master rate equations which give us the time
dependence of the temperature and fugacities \cite{LMW}. Initial values of the
temperature, quark fugacity, antiquark fugacity, and gluon fugacity of the
quark-gluon plasma are provided in Ref. \cite{XSYX}. Similar initial 
temperatures but different initial fugacities were obtained in Ref. \cite{LMW}
via free streaming.
Solving the master rate equations with the initial values, 
we obtain 0.64, 0.64, and 0.76 individually for the quark
fugacity, antiquark fugacity, and gluon fugacity at $T=1.01T_{\rm c}$
for central Au-Au collisions at $\sqrt {s_{NN}}=200$ GeV and 1, 1, and 1
for central Pb-Pb collisions at $\sqrt {s_{NN}}=5.5$ TeV. The corresponding
chemical potentials are -0.078, 0.078, and -0.048 for the quark, the
antiquark, and the gluon at the RHIC energy and 0, 0, and 0 at the LHC energy.
Therefore, in a volume $V$ of the quark-gluon plasma created in heavy ion
collisions from the highest RHIC energy to the highest LHC energy, 
the quark chemical potential is nearly or exactly zero, and we have 
\begin{equation}
\ln \Xi = \frac {4\pi g_Q V}{(2\pi)^3} \int_0^\infty d\mid \vec {p} \mid 
\vec {p}^2 \ln (1+e^{-\beta \sqrt {\vec {p}^2 +m^2}}). 
\end{equation}
With the variable $z=\beta \sqrt {\vec {p}^2 + m^2}$
the total energy of the quark system is
\begin{eqnarray}
E_q & = &  - \frac {\partial }{\partial \beta }\ln \Xi 
= \frac {4\pi g_Q V}{(2\pi)^3 \beta^4} \int_{m\beta}^\infty dzz^2
\sqrt {z^2 -m^2\beta^2}/(e^z+1)   \nonumber \\
& = & \frac {12\pi g_Q V}{(2\pi)^3\beta^4} \int_{m\beta}^\infty dzz
\sqrt {z^2 -m^2\beta^2} \ln (1+e^{-z})      \nonumber \\
& + & \frac {4\pi g_Q m^2V}{(2\pi)^3\beta^2} \int_{m\beta}^\infty dz
\frac {z\ln (1+e^{-z})}{\sqrt {z^2 -m^2\beta^2}} ,
\end{eqnarray}
and the pressure of the quark system is
\begin{eqnarray}
P_q & = & \frac {1}{\beta}\frac {\partial }{\partial V }\ln \Xi 
=\frac {4\pi g_Q}{(2\pi)^3\beta^4} \int_{m\beta}^\infty dzz
\sqrt {z^2 -m^2\beta^2} \ln (1+e^{-z})  \nonumber \\
& = & \frac {1}{3} \frac {E_q}{V}
- \frac {4\pi g_Q m^2}{3(2\pi)^3\beta^2} \int_{m\beta}^\infty dz
\frac {z\ln (1+e^{-z})}{\sqrt {z^2 -m^2\beta^2}}.
\end{eqnarray}
The quark entropy density is
\begin{equation}
s_q = \frac {1}{T} (\frac {E_q}{V}+P_q).
\end{equation}
The total energy, pressure and entropy density of the antiquark system are
written similarly. For a quark and an antiquark in the volume $V_{\rm c}$, 
their entropy is proportional to
\begin{equation}
S_{q\bar{q}} = 
s_q V_{\rm c}+s_{\bar{q}}V_{\rm c}.
\end{equation}

Below we show three examples about the system of quarks and antiquarks. 
In the first example we consider charm quarks
and charm antiquarks. If they are thermalized,
the above formulae can be applied with $g_Q=6$ for the charm quark and 
$g_{\bar Q}=6$ for the charm antiquark.
For the color singlet of the charm quark and the charm antiquark, 
$\frac {1}{3}$ is multiplied to $S_{c\bar c}$  to get the entropy of the color
singlet, and
$\frac{1}{3}TS_{c\bar c}=1.72 \times 10^{-3}$ GeV where $T=1.01T_{\rm c}$.
This indicates that the potential of the charm quark and the charm
antiquark almost equals the free energy. In the second example we consider
charm quarks and down antiquarks. For the color singlet of a charm
quark and a massless down antiquark (a down antiquark with a mass of 0.32 GeV
\cite{ZXG,ZX}), $\frac {1}{3}TS_{c\bar d}=0.0834$ GeV (0.0594 GeV). By
comparison with the free energy 0.123 GeV and the subtraction of the total 
kinetic energy of the quark and the antiquark from the internal energy, 
we realize that the free energy
can not be taken as the potential of the charm quark and the down antiquark.
In the third example we consider bottom (top) quarks and down antiquarks. 
For the color singlet of a bottom (top)
quark and a down antiquark with a mass of 0 or 0.32 GeV,
$\frac {1}{3}TS_{b\bar d}=0.0825$ GeV  or 0.0591 GeV
($\frac {1}{3}TS_{t\bar d}=0.0825$ GeV  or 0.0591 GeV).
We conclude that the free energy can not be taken as the potential of 
the bottom (top) quark and the down antiquark.

The grand partition function for a meson system is
\begin{equation}
\Xi = \prod\limits_l (1-e^{\beta (\mu - \varepsilon_l)})^{-\omega_l},
\end{equation}
where $\mu$ is the meson chemical potential, and $\omega_l$ is the number of
states corresponding to the meson energy $\varepsilon_l$. Coalescence of quarks
and antiquarks forms mesons at $T_{\rm c}$. Pions as the dominant species of
hadronic matter have the chemical potential close to zero from the highest
RHIC energy to the highest LHC energy.
In the relativistic case and at $\mu =0$,
\begin{equation}
\ln \Xi = -\frac {4\pi g_M V}{(2\pi)^3} 
\int_0^\infty d\mid \vec {p} \mid \vec {p}^2 \ln
(1-e^{-\beta \sqrt {\vec {p}^2 +m^2}}),
\end{equation}
where $g_M$ is the spin-isospin degeneracy factor. With the variable
$z=\beta \sqrt {\vec {p}^2+m^2}$
the total energy of the meson system is
\begin{eqnarray}
E_m & = &  - \frac {\partial }{\partial \beta }\ln \Xi 
= \frac {4\pi g_M V}{(2\pi)^3\beta^4} \int_{m\beta}^\infty dzz^2
\sqrt {z^2 -m^2\beta^2}/(e^z-1)   \nonumber \\
& = & -\frac {12\pi g_M V}{(2\pi)^3\beta^4} \int_{m\beta}^\infty dzz
\sqrt {z^2 -m^2\beta^2} \ln (1-e^{-z})      \nonumber \\
& & - \frac {4\pi g_M m^2 V}{(2\pi)^3\beta^2} \int_{m\beta}^\infty dz
\frac {z\ln (1-e^{-z})}{\sqrt {z^2 -m^2\beta^2}},
\end{eqnarray}
and the pressure of the meson system is
\begin{eqnarray}
P_m & = & \frac {1}{\beta}\frac {\partial }{\partial V }\ln \Xi 
=-\frac {4\pi g_M}{(2\pi)^3\beta^4} \int_{m\beta}^\infty dzz
\sqrt {z^2 -m^2\beta^2} \ln (1-e^{-z})     \nonumber \\
& = & \frac {1}{3} \frac {E_m}{V}
+ \frac {4\pi g_M m^2}{3(2\pi)^3\beta^2} \int_{m\beta}^\infty dz
\frac {z\ln (1-e^{-z})}{\sqrt {z^2 -m^2\beta^2}}.
\end{eqnarray}
The meson entropy density is
\begin{equation}
s_m = \frac {1}{T} (\frac {E_m}{V}+P_m),
\end{equation}
and the meson entropy in the volume $V_{\rm c}$ is proportional to
\begin{equation}
S_m = s_m V_{\rm c}.
\end{equation}
To get an impression on $TS_m$, we take the unrealistic case that $J/\psi$ is
thermalized in hadronic matter. With the $J/\psi$ mass 2.85 GeV/$c^2$ at
$T=0.99T_{\rm c}$ \cite{ZX}, we obtain $TS_{J/\psi}=1.98 \times 10^{-6}$ GeV
at $g_M=3$ while $S_{J/\psi}$ is the $J/\psi$ entropy in the volume 
$V_{\rm c}$. Therefore, 
the quark-antiquark potential in a $J/\psi$ meson equals the quark-antiquark
free energy. For the system of massless pions we have already obtained the
result that the free energy can be taken as the potential.
While the meson mass increases from 0, $E_m$, $P_m$, $s_m$, and
$TS_m$ decrease. Then, the result is also true in a meson with a nonzero mass.

While the quark chemical potential and the meson chemical potential are not
zero, they are negative \cite{LMW}, which corresponds to chemical
nonequilibrium of matter created in relativistic heavy ion collisions below
the highest RHIC energy. The grand partition
function for the quark system in Eq. (13) and the one for the
meson system in Eq. (19) are reduced. $E_q$, $P_q$, $s_q$, $S_{q\bar q}$, 
$E_m$, $P_m$, $s_m$, and $S_m$ decrease. 
Then, we can infer that the free energy can 
be taken as the potential in the meson system, the free energy can not be taken
as the potential of a heavy quark (antiquark) and a light antiquark (quark)
in the quark-gluon plasma, and the potential of a heavy quark and a heavy
antiquark almost equals the free energy.

Finally, we note that we can not use the entropy density obtained in lattice
calculations to deal with the relation between the quark-antiquark potential
and the quark-antiquark free energy. This entropy density includes not only
the one of quarks and antiquarks but also the one of
gluons. Nobody has separated the entropy density of quarks and antiquarks
from the entropy density obtained in lattice calculations.

\section{Summary}
The entropy of a particle system depends on the particle mass. When the 
system's
temperature is larger than the critical temperature, the product of the
temperature and the entropy of a massless quark-antiquark pair is
comparable to the quark-antiquark free energy, and the free energy can not 
be taken as the potential of a massless quark and a massless antiquark. 
This is also true while the quark or the antiquark is massless, but not while
both the quark and the antiquark are heavy. When the system's
temperature is smaller than the critical temperature, the product of the
temperature and a meson's entropy is small or negligible in comparison with 
the quark-antiquark free energy, and the free energy can be 
taken as the quark-antiquark potential to a good approximation, which is in
agreement with the result of the lattice calculations in Refs. \cite{RHS,BR}.

\vspace{0.5cm}
\leftline{\bf Acknowledgements}
\vspace{0.5cm}

This work was supported by the National Natural Science Foundation of China
under Grant No. 11175111.

\end{document}